\documentclass{icrc29_astro}
\usepackage{graphicx,amssymb,amsmath,times}
\newcommand{\SNe}{SNe}

\newcommand\etainj{\eta_\mathrm{inj}}
\newcommand{\fPL}{f_\mathrm{PL}(p)}
\newcommand{\pmax}{p_\mathrm{max}}

\newcommand{\RfsRcd}{R_\mathrm{FS}/R_\mathrm{CD}}
\newcommand{\RFS}{R_\mathrm{FS}}
\newcommand{\RCD}{R_\mathrm{CD}}
\newcommand{\CD}{contact discontinuity}

\newcommand{\syn}{synchrotron}

\newcommand{\pion}{pion-decay}

\newcommand{\NL}{nonlinear}

\newcommand{\rel}{relativistic}
\newcommand{\Rel}{Relativistic}
\newcommand{\nonrel}{non\-rel\-a\-tiv\-is\-tic}

\newcommand{\TP}{test-particle}

\newcount\listno
\def\List{\global\advance \listno by 1 {(\the\listno)}}

\newcount\listcno
\def\Listc{\global\advance \listcno by 1 
	{({\expandafter{\romannumeral\listcno})\,}}}
	\def\newlistc{\listcno=0}
\begin{document}
\title[Efficient Cosmic Ray Ion Production in SNRs]{Efficient Cosmic Ray
    Ion Production in Young Supernova Remnants} 
\author[Ellison \& Cassam-Chena{\"\i}] {Donald
    C. Ellison$^a$ and Gamil Cassam-Chena{\"\i}$^b$ \\ 
(a) Physics Dept., North Carolina State Univ., Raleigh, NC 27695-8202,
    U.S.A. \\
(b) Department of Physics and Astronomy, Rutgers University,
136 Frelinghuysen Rd, Piscataway NJ 08854-8019, U.S.A. } 
\presenter{Presenter: Don Ellison (don$\_$ellison@ncsu.edu), \
    usa-ellison-D-abs1-og22-oral}

\maketitle

\begin{abstract}
The strong shocks in young supernova remnants (SNRs) should accelerate cosmic rays
(CRs) and no doubt exists that \rel\ electrons are produced in SNRs. However, direct
evidence that SNRs produce CR nuclei depends on seeing an unambiguous \pion\ feature
and this has not yet been obtained.  Nevertheless, the lack of an observed \pion\
feature does not necessarily mean that CR ions are not abundantly produced since ions
do not radiate efficiently.
If CR ions are produced efficiently by diffusive shock acceleration (DSA), their
presence will modify the hydrodynamics of the SNR and produce morphological effects
which can be clearly seen in radiation produced by electrons.
\end{abstract}

\vskip-30pt\hbox{}
\section{Introduction} \vskip-6pt
Particle acceleration influences the SNR evolution because the production of CRs
changes the effective equation of state of the shocked gas.  \Rel\ particles produce
less pressure for a given energy density than do \nonrel\ ones and high energy
particles can escape from the shock and carry away energy and pressure.
The softer
effective equation of state means that compression ratios well in excess of four can
be produced in non-radiative, collisionless shocks (e.g., \cite{JE91}).
An important morphological aspect of this CR-hydro coupling is that the
ratio of the forward shock radius, $\RFS$, to the radius of the \CD, $\RCD$, may be
much less than in the \TP\ case (e.g., \cite{DEB2000}\cite{EDB2004}).
If, as is generally believed, shocks put far more energy into accelerated ions than
electrons, it is the efficient production of cosmic ray {\it ions} that reduces
$\RfsRcd$ from \TP\ values. 
Therefore, measuring $\RfsRcd$ with modern X-ray telescopes may reveal the presence
of these otherwise unseen \rel\ ions.

\vskip-24pt\hbox{}
\section{Model and Results} \vskip-6pt
Full details of the model we use here are given in \cite{EC05} and references
therein. Briefly, we calculate the hydrodynamic evolution of a SNR coupled to
efficient DSA. The diffusive shock acceleration process is modeled with the algebraic
model of \cite{BE99} and \cite{EBB2000} where the injection efficiency is
parameterized and the superthermal spectrum, $f(p)$, is a broken power law, $\fPL$,
with an exponential turnover at high momenta, $f(p) \propto \fPL \exp{(-p/\pmax)}$.
The algebraic model solves the \NL\ DSA problem at each time step of the hydro
simulation given the shock speed, shock radius, ambient density and temperature, and
ambient magnetic field determined in the simulation. With the accelerated
distribution, an effective ratio of specific heats is calculated and used in the
hydrodynamic equations, completing the coupling between the two.

The injection parameter,
$\etainj$, is the fraction of total protons injected into the DSA process and values
$\etainj \gtrsim 10^{-4}$ typically yield efficient particle acceleration rates where
$10 \%$ to $99\%$ of the available energy flux goes into \rel\ protons.

As the forward shock (FS) overtakes fresh ambient medium material, the shock
accelerates these particles and produces a nonthermal distribution.\footnote{We
ignore pre-existing CRs and acceleration at the reverse shock and 
accelerate only thermal particles overtaken by the FS.}
Once the particle distribution is produced in a shell of material at the shock, it is
assumed to remain in that shell as the shell convects and evolves behind the shock.
During the evolution, particles experience adiabatic and \syn\ losses and 
in calculating the \syn\ emission and losses, we evolve the magnetic field as
described, for example, in \cite{Reynolds98}.

\begin{figure}[h]
\begin{center}
\includegraphics*[width=0.7\textwidth,angle=0,clip]{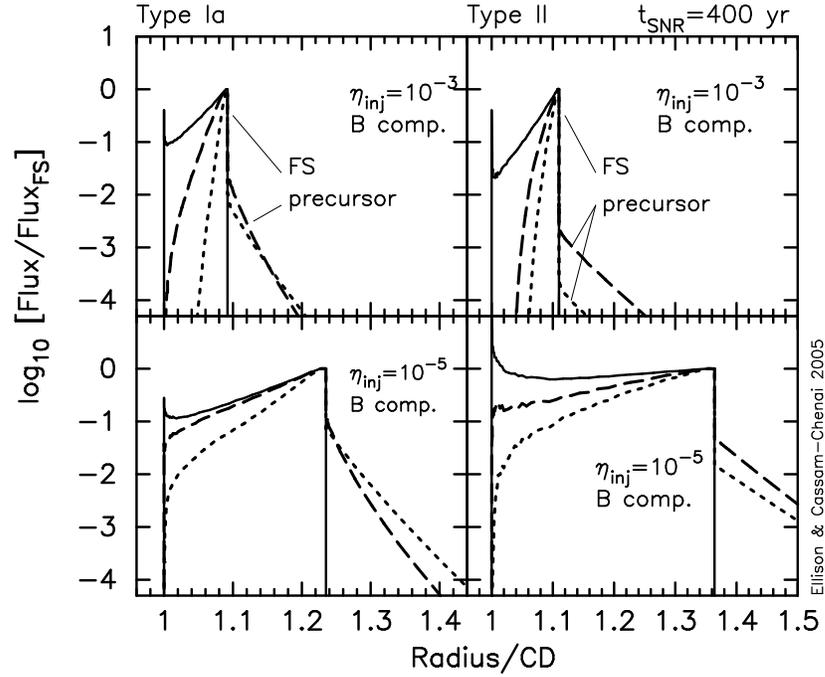}
\caption{\label {fig1} Radial \syn\ emission in three energy bands for two shock
  injection efficiencies, $\etainj$. The left two panels use typical type Ia
  parameters where the right two panels use type II parameters (for a complete list
  of parameters, see \cite{EC05}).  In all panels, the solid curve is radio (1--1.4
  GHz), the dashed curve is low energy X-rays (0.1--1 keV), and the dotted curve is
  high energy X-rays (1--10 keV). The flux of each band is normalized to its value at
  the FS. See \cite{EC05} for full details.}
\end{center}
\end{figure}

\newlistc

In Fig.~\ref{fig1} we plot the \syn\ emission as a function of radius for one radio
(1-1.4 GHz; solid curves) and two X-ray bands (0.1-1 keV dashed curves; 1-10 keV
dotted curves).  The two left-hand panels show results with strong CR production
($\etainj=10^{-3}$) and \TP\ production ($\etainj=10^{-5}$) using typical type Ia
parameters. The two right-hand panels use typical type II parameters. We note: 
\Listc
In the interaction region between the contact discontinuity and the
forward shock, the X-ray \syn\ falls off more rapidly than the radio emission due to
strong \syn\ losses.
\Listc
With efficient DSA (top panels), the X-ray fall-off is extremely rapid
and the X-ray emission can appear as an extremely thin sheet at the FS.
\Listc
The precursor outside of the FS drops sharply due to the compressed
field, making the X-ray precursor faint and difficult to detect compared to the
emission at the FS.
\Listc
Comparing the $\etainj=10^{-3}$ panels against the $\etainj=10^{-5}$ panels shows
that the distance between the CD and the FS is nearly a factor of two greater in the
test-particle case than with efficient DSA.  Since the limit of the shocked ejecta
gives an idea of the position of the CD, $\RfsRcd$ is measurable in several young
SNRs with {\it Chandra} and {\it XMM-Newton}, making this morphological difference a
powerful diagnostic for efficient DSA.

\begin{figure}[h]
\begin{center}
\includegraphics*[width=0.4\textwidth,angle=0,clip]{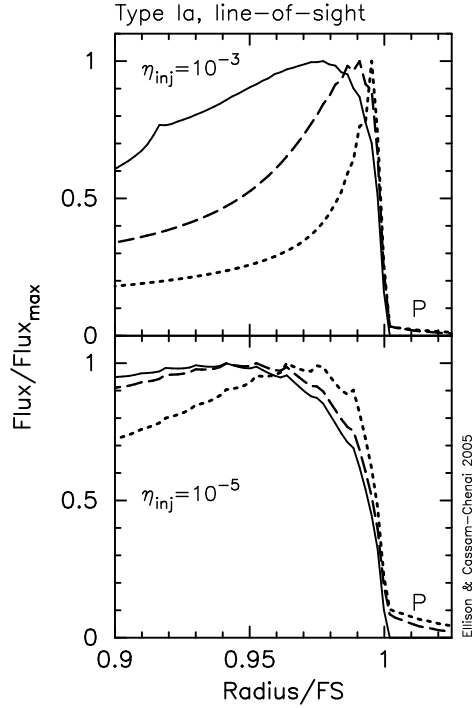}
\caption{\label {fig2} Line-of-sight projections for the radial distributions shown
  in Fig.~\ref{fig1} for type Ia normalized to the forward shock radius. As in
  Fig.~\ref{fig1}, the solid curve is radio (1--1.4 GHz), the dashed curve is low
  energy X-rays (0.1--1 keV), and the dotted curve is high energy X-rays (1--10 keV).
  Note that the radio emission (solid curves) peaks well within the X-ray emission.}
\end{center}
\end{figure}

In Fig.~\ref{fig2} we show line-of-sight (LOS) projections for the type Ia examples
of Fig.~\ref{fig1} plotted as a fraction of the FS radius. Note that in the LOS
projections the radio and X-ray peaks are offset at the FS: with or without efficient
DSA, the radio peak (solid curve) occurs inside the X-ray peaks.  Behavior such as
this is observed in several SNRs including G347 \cite{Lazendic2004}, Kepler
\cite{DeLaneyEtal2002}, Tycho \cite{DecourchelleEtal2001}, and Cas A
\cite{LongEtal2003}.
Note also that because of projection effects, the maximum emission occurs inside of
the FS.

\begin{figure}[h]
\begin{center}
\includegraphics*[width=0.45\textwidth,angle=0,clip]{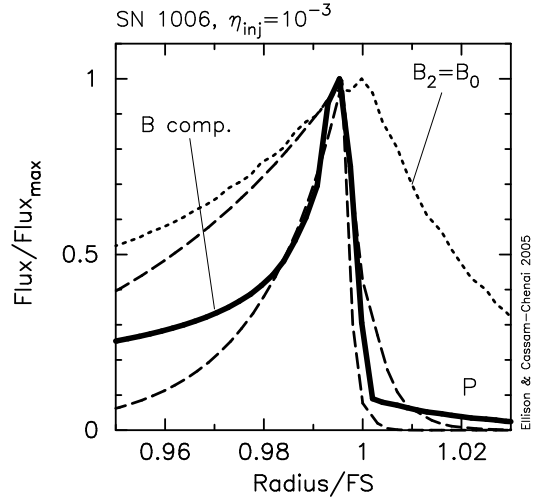}
\caption{\label {fig3} Comparison of X-ray LOS profiles from the CR-Hydro model with
{\it Chandra} observations of SN 1006. The dashed curves roughly span the maximum and
minimum scale heights determined by \cite{BambaEtal2003} where they assumed
exponential profiles.  Using their Table~4, we set the maximum (minimum) upstream
scale height to be 3 (1) arcsec, and the downstream maximum (minimum) scale height to
be 30 (10) arcsec (the radius of SN 1006 is about $0.25^\circ$).  The solid curve is
the X-ray emission in the 1.2--2 keV band using our compressed $B$ model and for
comparison, we show (dotted curve) the 1.2--2 keV band without compressing the
field. We have positioned the peaks of the dashed curves to match the solid curve. }
\end{center}
\end{figure}

In Fig.~\ref{fig3} we compare our type Ia prototype model with $\etainj = 10^{-3}$
to SN 1006 observations represented with dashed lines which roughly span
the maximum and minimum scale heights determined by
\cite{BambaEtal2003} (see their Table~4). 
As emphasized by \cite{BKV2003}, the shortest scale heights occur inside the forward
shock and are produced by projection effects when $B$ is compressed and there is a
sharp drop in emissivity behind the shock. The actual upstream precursor (indicated
with a ``P'' in Fig.~\ref{fig3}) has a much longer scale height as expected from TeV
electrons but is not easily discernable with {\it Chandra} against background
emission.
Our efficient acceleration model fits the data quite well, but a \TP\ model (not
shown), clearly does not.  Our results, in agreement with those of \cite{BKV2003},
provide evidence for highly compressed magnetic fields and efficient DSA.

\vskip-24pt\hbox{}
\section{Discussion and Conclusions} \vskip-6pt
An important signature of efficient CR ion production is a large reduction in the
ratio of the radius of the forward shock to the radius of the \CD, $\RfsRcd$, caused
by compression ratios $\gg 4$.
This effect may explain observations of $\RfsRcd \sim 1$ in Tycho's
and Kepler's SNRs. Type II \SNe\ with pre-SN winds may experience efficient DSA yet
still show large $\RfsRcd \sim 1.3$--$1.4$, consistent with observations of Cas A and
1E0102.2-7219.
Another sign of efficient DSA is the presence of short scale heights seen in
nonthermal X-ray emission. Short scale heights are predicted
because the shock will strongly compress the downstream magnetic field and \syn\
losses will lower the emissivity immediately behind the FS. 
Also, projection effects should result in the distinct separation of the radio and
X-ray peaks.

The short scale heights seen in SN 1006 \cite{BambaEtal2003}, are most naturally
explained as sharply peaked emission behind the FS seen in projection (e.g.,
\cite{BKV2003}).  The actual upstream precursor has a long scale length, as expected
for TeV electrons, but is weak enough to avoid detection.  Supernova remnant SN 1006
seems a clear case where the efficient production of CR ions is taking place, but
remnants such as Tycho's and Kepler's, with $\RfsRcd \sim 1$, are also likely
candidates.

{\bf Acknowledgements:} D.C.E. wishes to acknowledge support from a NASA grant
(ATP02-0042-0006). \vskip-12pt

\newcommand\itt{ }
\newcommand\bff{ }
\newcommand{\aaDE}[3]{ 19#1, A\&A, #2, #3}
\newcommand{\aatwoDE}[3]{ 20#1, A\&A, #2, #3}
\newcommand{\aatwopress}[1]{ 20#1, A\&A, in press}
\newcommand{\aasupDE}[3]{ 19#1, {\itt A\&AS,} {\bff #2}, #3}
\newcommand{\ajDE}[3]{ 19#1, {\itt AJ,} {\bff #2}, #3}
\newcommand{\anngeophysDE}[3]{ 19#1, {\itt Ann. Geophys.,} {\bff #2}, #3}
\newcommand{\anngeophysicDE}[3]{ 19#1, {\itt Ann. Geophysicae,} {\bff #2}, #3}
\newcommand{\annrevDE}[3]{ 19#1, {\itt Ann. Rev. Astr. Ap.,} {\bff #2}, #3}
\newcommand{\apjDE}[3]{ 19#1, {\itt ApJ,} {\bff #2}, #3}
\newcommand{\apjtwoDE}[3]{ 20#1, {\itt ApJ,} {\bff #2}, #3}
\newcommand{\apjletDE}[3]{ 19#1, {\itt ApJ,} {\bff  #2}, #3}
\newcommand{\apjlettwoDE}[3]{ 20#1, {\itt ApJ,} {\bff  #2}, #3}
\newcommand{\apjpress}[1]{{\itt #1, ApJ,} in press}
\newcommand{\apjletpress}{{\itt ApJ(Letts),} in press}
\newcommand{\apjsDE}[3]{ 19#1, {\itt ApJS,} {\bff #2}, #3}
\newcommand{\apjstwoDE}[3]{ 19#1, {\itt ApJS,} {\bff #2}, #3}
\newcommand{\apjsubDE}[1]{ 19#1, {\itt ApJ}, submitted.}
\newcommand{\apjsubtwoDE}[1]{ 20#1, {\itt ApJ}, submitted.}
\newcommand{\appDE}[3]{ 19#1, {\itt Astropart. Phys.,} {\bff #2}, #3}
\newcommand{\apptwoDE}[3]{ 20#1, {\itt Astropart. Phys.,} {\bff #2}, #3}
\newcommand{\araaDE}[3]{ 19#1, {\itt ARA\&A,} {\bff #2},
   #3}
\newcommand{\assDE}[3]{ 19#1, {\itt Astr. Sp. Sci.,} {\bff #2}, #3}
\newcommand{\grlDE}[3]{ 19#1, {\itt G.R.L., } {\bff #2}, #3} 
\newcommand{\icrcplovdiv}[2]{ 1977, in {\itt Proc. 15th ICRC (Plovdiv)},
   {\bff #1}, #2}
\newcommand{\icrcsaltlake}[2]{ 1999, {\itt Proc. 26th Int. Cosmic Ray Conf.
    (Salt Lake City),} {\bff #1}, #2}
\newcommand{\icrcsaltlakepress}[2]{ 19#1, {\itt Proc. 26th Int. Cosmic Ray Conf.
    (Salt Lake City),} paper #2}
\newcommand{\icrchamburg}[2]{ 2001, {\itt Proc. 27th Int. Cosmic Ray Conf.
    (Hamburg),} {\bff #1}, #2}
\newcommand{\JETPDE}[3]{ 19#1, {\itt JETP, } {\bff #2}, #3}
\newcommand{\jgrDE}[3]{ 19#1, {\itt J.G.R., } {\bff #2}, #3}
\newcommand{\mnrasDE}[3]{ 19#1, {\itt MNRAS,} {\bff #2}, #3}
\newcommand{\mnrastwoDE}[3]{ 20#1, {\itt MNRAS,} {\bff #2}, #3}
\newcommand{\mnraspress}[1]{ 20#1, {\itt MNRAS,} in press}
\newcommand{\natureDE}[3]{ 19#1, {\itt Nature,} {\bff #2}, #3}
\newcommand{\naturetwoDE}[3]{ 20#1, {\itt Nature,} {\bff #2}, #3}
\newcommand{\nucphysA}[3]{#1, {\itt Nuclear Phys. A,} {\bff #2}, #3}
\newcommand{\pfDE}[3]{ 19#1, {\itt Phys. Fluids,} {\bff #2}, #3}
\newcommand{\phyreptsDE}[3]{ 19#1, {\itt Phys. Repts.,} {\bff #2}, #3}
\newcommand{\physrevEDE}[3]{ 19#1, {\itt Phys. Rev. E,} {\bff #2}, #3}
\newcommand{\prlDE}[3]{ 19#1, {\itt Phys. Rev. Letts,} {\bff #2}, #3}
\newcommand{\prltwoDE}[3]{ 20#1, {\itt Phys. Rev. Letts,} {\bff #2}, #3}
\newcommand{\revgeospphyDE}[3]{ 19#1, {\itt Rev. Geophys and Sp. Phys.,}
   {\bff #2}, #3}
\newcommand{\rppDE}[3]{ 19#1, {\itt Rep. Prog. Phys.,} {\bff #2}, #3}
\newcommand{\rpptwoDE}[3]{ 20#1, {\itt Rep. Prog. Phys.,} {\bff #2}, #3}
\newcommand{\ssrDE}[3]{ 19#1, {\itt Space Sci. Rev.,} {\bff #2}, #3}
\newcommand{\ssrtwoDE}[3]{ 20#1, {\itt Space Sci. Rev.,} {\bff #2}, #3}
\newcommand{\scienceDE}[3]{ 19#1, {\itt Science,} {\bff #2}, #3} 
\newcommand{\spDE}[3]{ 19#1, {\itt Solar Phys.,} {\bff #2}, #3} 
\newcommand{\spuDE}[3]{ 19#1, {\itt Sov. Phys. Usp.,} {\bff #2}, #3}

\end{document}